\documentclass[12pt,thmsa]{article}
\usepackage{amsfonts}

\usepackage{sw20lart}

\input{tcilatex}
\input tcilatex
\begin{document}

\title{The Modal Interpretation of Algebraic Quantum Field Theory}
\author{Rob Clifton \\
Department of Philosophy, 1001 Cathedral of Learning, \\
University of Pittsburgh, Pittsburgh, PA 15260\\
(E-mail: rclifton+@pitt.edu)}
\maketitle

\begin{abstract}
In a recent Letter, Dieks \cite{dieks0} has proposed a way to implement the
modal interpretation of quantum theory in algebraic quantum field theory. We
show that his proposal fails to yield a well-defined prescription for which
observables in a local spacetime region possess definite values. On the
other hand, we demonstrate that there \textit{is} a well-defined and \textit{%
unique} way of extending the modal interpretation to the local algebras of
quantum field theory. This extension, however, faces a potentially serious
difficulty in connection with ergodic states of a field.


\end{abstract}

\section{The Modal Interpretation of Nonrelativistic Quantum Theory}

The modal interpretation is actually a family of interpretations sharing the
feature that a system's density operator constrains the possibilities for
assigning definite values to its observables \cite{healeybook}--\cite
{modalbook}. We start with a brief review of the basic idea behind these
interpretations from an algebraic point of view \cite{zc}--\cite{hc}.

Consider a `universe' comprised of a quantum system $U$, with finitely many
degrees of freedom, represented by a Hilbert space \textsf{H}$%
_{U}=\bigotimes_{i}\mathsf{H}_{i}\,$. At any given time, $U$ will occupy a
pure vector state $x\in \mathsf{H}_{U}$ that determines a reduced density
operator $D_{S}$ on the Hilbert space \textsf{H}$_{S}=\bigotimes_{i\in S}%
\mathsf{H}_{i}$ of any subsystem $S\subseteq U$. Let $\mathcal{B}(\mathsf{H}%
_{S})$ denote the algebra of all bounded operators on \textsf{H}$_{S}$, and
for any single operator or family of operators $T\subseteq \mathcal{B}(%
\mathsf{H}_{S})$, let $T^{\prime }$ denote its commutant (i.e., all
operators on $\mathsf{H}_{S}$ that commute with those in $T$). Let $P_{S}$
denote the projection onto the range of $D_{S}$, and consider the subalgebra
of $\mathcal{B}(\mathsf{H}_{S})$ given by the direct sum 
\begin{equation}
\mathcal{M}_{S}\equiv P_{S}^{\perp }\mathcal{B}(\mathsf{H}_{S})P_{S}^{\perp
}+D_{S}^{\prime \prime }P_{S}.  \label{MS}
\end{equation}
When $S$ is not entangled with its environment $\overline{S}$ (represented
by \textsf{H}$_{\overline{S}}=\bigotimes_{i\notin S}$\textsf{H}$_{i}$), $%
D_{S}$ will itself be a pure state, induced by a unit vector $y\in \mathsf{H}%
_{S}$. In this case, $\mathcal{M}_{S}$ consists of all operators with $y$ as
an eigenvector --- the self-adjoint members of which are taken, in orthodox
quantum theory, to be the observables of $S$ with definite values. On the
other hand, when there \textit{is} entanglement and $D_{S}$ is mixed --- in
particular, in the extreme case $P_{S}=I$ (where essentially every vector
state $y\in $ \textsf{H}$_{S}$ is a component of the mixture) --- then $%
\mathcal{M}_{S}\ =D_{S}^{\prime \prime }$, and $\mathcal{M}_{S}$ consists
simply of all functions of $D_{S}$. In this case, orthodox quantum theory
has nothing to say about the properties of $S$. Thus, when $S$ is
Schr\"{o}dinger's cat entangled with some potentially cat killing device $%
\overline{S}$, we get the infamous measurement problem.

In non-atomic versions of the modal interpretation (\cite{vFbook}, \cite
{kochen}--\cite{clifton2}), there is no preferred partition of the universe
into subsystems. Any particular subsystem $S\subseteq U$ is taken to have
definite values for all the self-adjoint operators that lie in \thinspace $%
\mathcal{M}_{S}$, and this applies \textit{whether or not} $D_{S}$ is pure.
The values of these observables are taken to be distributed according to the
usual Born rule. Thus, the expectation of any observable $A\in \mathcal{M}%
_{S}$ is $\limfunc{Tr}(D_{S}A)$, and the probability that $A$ possesses some
particular value $a_{j}$ is $\limfunc{ Tr}(D_{S}P^{j})$, where $P^{j}\in 
\mathcal{M}_{S}$ is the corresponding eigenprojection of $A$. As in orthodox
quantum theory, \textit{which} precise value for $A$ occurs on a given
occasion (from amongst those with nonzero probability in state $D_{S}$) is
not fixed by the interpretation. However, the occurrence of a value does not
require that it be `measured'. And, unlike orthodox quantum theory, no
miraculous collapse is needed to solve the measurement problem. Instead,
after a typical unitary `measurement' interaction between two parts of the
universe --- $O$ the `measured' system, and $A$ the apparatus ---
decoherence induced by $A$'s coupling to the environment $\overline{O\cup A}$
will force the density operator $D_{A}$ of the apparatus to diagonalize in a
basis extremely close to one which diagonalizes the pointer observable of $A$
\cite{guidomeir}. Thus, after the measurement, the definite-valued
observables in $\mathcal{M}_{A}$ will be such that the pointer points!

In atomic versions of the modal interpretation (\cite{healeybook}, \cite{BD}%
--\cite{toronto}), one does not tell a separate story about the
definite-valued observables for each subsystem $S\subseteq U$. Rather, $S$
is taken to inherit properties from those of its atomic components,
represented by the individual Hilbert spaces in $\bigotimes_{i\in S}\mathsf{H%
}_{i}$. In the approach favoured by Dieks \cite{diekshealey} (cf. \cite{BD}%
), each atomic system $i$ possesses definite values for all the observables
in $\mathcal{M}_{i}$, as determined by the corresponding atomic density
operator $D_{i}$ in accordance with (\ref{MS}). The definite-valued
observables of $S$ are then built up by embedding each $\mathcal{M}_{i}$ in $%
\mathcal{B}(\mathsf{H}_{S})$ (via tensoring it with the identity on \textsf{H%
}$_{\overline{S}}$), and taking the von Neumann subalgebra of $\mathcal{B}(%
\mathsf{H}_{S})$ generated by all these embeddings. The definite properties
of $S$ will therefore include all projections $\bigotimes_{i\in
S}P_{i}^{j(i)}$ that are tensor products of the spectral projections of the
individual atomic density operators, and their joint probabilities are again
taken to be given by the usual Born rule $\limfunc{Tr}[D_{S}(\bigotimes_{i%
\in S}P_{i}^{j(i)})]$. In the presence of decoherence, the expectation is
that atomic versions of the modal interpretation can yield essentially the
same resolution of the measurement problem as do non-atomic versions \cite
{diekshealey,vermaas}.

In both versions of the modal interpretation, the observables with definite
values must change over time as a function of the (generally, non-unitary)
evolution of the reduced density operators of the systems involved. In
principle, this evolution can be determined from the (unitary)
Schr\"{o}dinger evolution of the universal state vector $x\in $ \textsf{H}$%
_{U}$ \cite{tough}. But the evolution of the precise values of the
definite-valued observables themselves is not determined by the
Schr\"{o}dinger equation. Various more or less natural proposals have been
made for `completing' the modal interpretation with a dynamics for values (%
\cite{BD},\cite{vermaas0}--\cite{Ax-Kochen}). Unfortunately, it has been
shown that the most natural proposals for a dynamics, particularly in the
case of atomic modal interpretations, must break Lorentz-invariance \cite{cd}%
. Most of Dieks' recent Letter \cite{dieks0} is concerned to address this
dynamics problem by appropriating ideas from the decoherent histories
approach to quantum theory (cf. \cite{meir}). However, we shall focus here
entirely on the viability of Dieks' new proposal for picking out
definite-valued observables of a relativistic quantum field that are
associated with approximately point-sized regions of Minkowski spacetime (%
\cite{dieks0}, Sec. 5).

\section{Critique of Dieks' Proposal}

Dieks' stated aim is to see if the modal interpretation can achieve sensible
results in the context of quantum field theory. For this purpose, he adopts
the formalism of algebraic quantum field theory because of its generality 
\cite{haag}--\cite{russians}. In the concrete `Haag-Araki' approach, one
supposes that a quantum field on Minkowski spacetime $M$ will associate to
each bounded open region $O\subseteq M$ a von Neumann algebra $\mathcal{R}%
(O) $ of observables measurable in that region, where the collection $\{%
\mathcal{R}(O):O\subseteq M\}$ acts irreducibly on some fixed Hilbert space 
\textsf{H}. It is then natural to treat each open region $O$ and associated
algebra $\mathcal{R}(O)$ as a quantum system in its own right. Given any
(normal) state $\rho $ of the field (where $\rho $ is a state functional on $%
\mathcal{B}(\mathsf{H})$), we can then ask which observables in $\mathcal{R}%
(O)$ are picked out as definite-valued by the restriction, $\rho {}_{O}$, of
the state $\rho $ to $\mathcal{R}(O)$.

The difficulty for the modal interpretation (though Dieks himself does not
put it this way) is that when $O$ has nonempty spacelike complement $%
O^{\prime }$, $\mathcal{R}(O)$ will typically be a type III factor that
contains no nonzero finite projections (\cite{haag}, Sec. V.6; \cite
{russians}, Sec. 17.2). Because of this, $\mathcal{R}(O)$ cannot contain
compact operators, like density operators, all of whose (non-null) spectral
projections are finite-dimensional. As a result, there is no density
operator in $\mathcal{R}(O)$ that can represent $\rho {}_{O}$. Moreover, if
we try to apply the standard modal prescription based on Eq. (\ref{MS}) to a
density operator in $\mathcal{B}(\mathsf{H})$ that agrees with $\rho
{}_{O}{} $, there is no guarantee that the resulting set of observables will
pick out a subalgebra of $\mathcal{R}(O)$, and we will be left with nothing
to say about which observables have definite values in $O$. The moral Dieks
draws from this is that ``We can therefore not take the open spacetime
regions and their algebras as fundamental, if we want an interpretation in
terms of (more or less) localized systems whose properties would specify an
event'' (\cite{dieks0}, p. 322). We shall see in the next section that this
conclusion is overly pessimistic.

In any case, Dieks' strategy for dealing with the problem is to exploit the
fact that, in most models of the axioms of algebraic quantum field theory,
the local algebras associated with diamond shaped spacetime regions (i.e.,
regions given by interior of the intersection of the causal future and past
of two spacetime points) have the split property. The property is that for
any two concentric diamond shaped regions $\diamondsuit _{r}$ $,\diamondsuit
_{r+\epsilon }$ $\subseteq $ $M$, with radii $r$ and $r+\epsilon $, there is
a type I `interpolating' factor $\mathcal{N}_{r+\epsilon }$ such that $%
\mathcal{R}(\diamondsuit _{r})\subset \mathcal{N}_{r+\epsilon }\subset 
\mathcal{R}(\diamondsuit _{r+\epsilon })$. Now since $\mathcal{N}%
_{r+\epsilon }\approx $ $\mathcal{B}(\overline{\mathsf{H}})$ for some
Hilbert space $\overline{\mathsf{H}}$, we know there is always a unique
density operator $D_{r+\epsilon }\in \mathcal{B}(\overline{\mathsf{H}})$
that agrees with $\rho \ $on $\mathcal{N}_{r+\epsilon }$, and therefore with 
$\rho _{\diamondsuit _{r}}$. The proposal is, then, to take both $r$ and $%
\epsilon $ to be fixed small numbers and apply the prescription in Eq. (\ref
{MS}) to this density operator $D_{r+\epsilon }$, yielding a definite-valued
subalgebra $\mathcal{M}_{r+\epsilon }\subseteq \mathcal{N}_{r+\epsilon }$.
This, according to Dieks, should give an approximation indication of which
observables have definite values at the common origin of the two diamonds in
``the classical limiting situation in which classical field and particle
concepts become approximately applicable'' (\cite{dieks0}, p. 323). Thus,
Dieks proposes to build up an atomic modal interpretation of the field as
follows. (i) We subdivide (approximately) the whole of spacetime $M$ into a
collection of non-overlapping diamond regions $\diamondsuit _{r}$, with some
fixed small radius $r$. (ii) We choose some fixed small $\epsilon $ and an
interpolating factor $\mathcal{N}_{r+\epsilon }$ for each diamond, using its 
$\rho $-induced density operator $D_{r+\epsilon }$ and Eq. (\ref{MS}) to
determine the definite-valued observables $\mathcal{M}_{r+\epsilon }$ $%
\subset \mathcal{R}(\diamondsuit _{r+\epsilon })$ to be loosely associated
with the origin. (iii) Finally, we build up definite-valued observables
associated with collections of diamonds, and define their joint
probabilities in the usual way via Born's rule (defining transition
probabilities between values of observables associated with
timelike-separated diamonds using the familiar multi-time generalization of
that rule employed in the decoherent histories approach).

As things stand, there is much arbitrariness in this proposal that enters
into the stages (i) and (ii). Dieks himself recognizes the arbitrariness in
the size of the partition of $M$ chosen. He also acknowledges that this
arbitrariness cannot be eliminated by passing to the limit $r,\epsilon
\rightarrow 0$, because the intersection of the algebras associated with any
collection of concentric diamonds is always the trivial algebra $\frak{C}I$ 
\cite{wightman}. Indeed, one would have thought that this undermines any
attempt to formulate an \textit{atomic} modal interpretation in this
context, because it forces the choice of `atomic diamonds' in the partition
of $M$ to be essentially arbitrary. Dieks appears to suggest that this
arbitrariness will become unimportant in some classical limit of
relativistic quantum field theory in which we should recover ``the classical
picture according to which field values are attached to spacetime points'' (%
\cite{dieks0}, p. 325). But it is not sufficient for the success of a
proposal for interpreting a relativistic \textit{quantum }field theory that
the interpretation give sensible results in the limit of \textit{classical}
relativistic (or nonrelativistic) field theory. Indeed, the only relevant
limit would appear to be the nonrelativistic limit; i.e., Galilean quantum
field theory. But, there, one still needs to spatially smear
``operator-valued'' fields at each point to obtain a well-defined algebra of
observables in a spatial region \cite{wreck}, so there will again be no
natural choice to make for atomic spatial regions or algebras.

There is also another, more troubling, degree of arbitrariness at step (ii)
in the choice of the type I interpolating factor $\mathcal{N}_{r+\epsilon }$
about each origin point. For any fixed partition and fixed $r,\epsilon >0$,
we can always sub-divide the interval $(r,r+\epsilon )$ further, and then
the split property implies the existence of a pair of interpolating type I
factors satisfying

\begin{equation}
\mathcal{R}(\diamondsuit _{r})\subset \mathcal{N}_{r+\epsilon /2}\subset 
\mathcal{R}(\diamondsuit _{r+\epsilon /2})\subset \mathcal{N}_{r+\epsilon
}\subset \mathcal{R}(\diamondsuit _{r+\epsilon }).  \label{hans}
\end{equation}
The problem is that we now face a nontrivial choice deciding which of these
factors' $\rho $-induced density operators to use to pick out the
definite-valued observables in the state $\rho $ associated with the origin.
If we pick $D_{r+\epsilon }\in \mathcal{N}_{r+\epsilon }$, then by (\ref{MS}%
) all observables $A\in \mathcal{N}_{r+\epsilon }$ that share the same
spectral projections as $D_{r+\epsilon }$ will have definite values.
However, \textit{no} such $A$ can lie in $\mathcal{M}_{r+\epsilon
/2}\subseteq \mathcal{N}_{r+\epsilon /2}$, nor even in $\mathcal{R}%
(\diamondsuit _{r+\epsilon /2})$. The reason is that $A$'s spectral
projections are \textit{finite in} $\mathcal{N}_{r+\epsilon }$. So if those
projections were also in the type III algebra $\mathcal{R}(\diamondsuit
_{r+\epsilon /2})$, they would have to be infinite in $\mathcal{R}%
(\diamondsuit _{r+\epsilon /2})$, and therefore also infinite projections in 
$\mathcal{N}_{r+\epsilon }$ --- which is impossible.

Clearly we can sub-divide the interval $(r,r+\epsilon )$ arbitrarily many
times in this way and obtain a monotonically decreasing sequence of type I
factors satisfying $\mathcal{R}(\diamondsuit _{r+\epsilon /2^{n+1}})\subset 
\mathcal{N}_{r+\epsilon /2^{n}}\subset \mathcal{R}(\diamondsuit _{r+\epsilon
/2^{n}})$ that all interpolate between $\mathcal{R}(\diamondsuit _{r})$ and $%
\mathcal{R}(\diamondsuit _{r+\epsilon })$. The sequence $\{\mathcal{N}%
_{r+\epsilon /2^{n}}\}_{n=0}^{\infty }$ has no least member, and its
greatest member, $\mathcal{N}_{r+\epsilon }$, is arbitrary, because we could
also further sub-divide the interval $(r+\epsilon /2,r+\epsilon )$ ad
infinitum. Thus there is no natural choice of interpolating factor for
picking out the observables definite at the origin, even if we restrict
ourselves to a `nice' decreasing sequence of interpolating factors of the
form $\{\mathcal{N}_{r+\epsilon /2^{n}}\}_{n=0}^{\infty }$.

On the other hand, it can actually be shown that $\mathcal{R}(\diamondsuit
_{r})=\bigcap_{n=0}^{\infty }\mathcal{N}_{r+\epsilon /2^{n}}$ (\cite{horuzy}%
, pp. 12-3; \cite{russians}, p. 426). Furthermore, suppose $\{\mathcal{N}%
_{r+\epsilon _{n}}\}_{n=0}^{\infty }$ is \textit{any} \textit{other}
decreasing type I sequence satisfying 
\begin{equation}
\mathcal{R}(\diamondsuit _{r+\epsilon _{n+1}})\subset \mathcal{N}%
_{r+\epsilon _{n}}\subset \mathcal{R}(\diamondsuit _{r+\epsilon
_{n}}),\;\epsilon _{0}=\epsilon ,\;\epsilon _{n}>\epsilon _{n+1},\;\lim
\epsilon _{n}=0.
\end{equation}
Then, since for any $n$ there will be a sufficiently large $n^{\prime }$
such that $\mathcal{N}_{r+\epsilon _{n^{\prime }}}\subset \mathcal{N}%
_{r+\epsilon /2^{n}}$ (and vice-versa), clearly $\mathcal{R}(\diamondsuit
_{r})=\bigcap_{n=0}^{\infty }\mathcal{N}_{r+\epsilon _{n}}$, and this
intersection will also be independent of $\epsilon $. It would seem, then,
that the natural way to avoid choosing between the myriad type I factors
that interpolate between $\mathcal{R}(\diamondsuit _{r})$ and $\mathcal{R}%
(\diamondsuit _{r+\epsilon })$ is to take the observables definite-valued at
the origin to be those in the intersection $\bigcap_{n=0}^{\infty }\mathcal{M%
}_{r+\epsilon _{n}}\subseteq \mathcal{R}(\diamondsuit _{r})$ (where, as
before, $\mathcal{M}_{r+\epsilon _{n}}$ is the modal subalgebra of $\mathcal{%
N}_{r+\epsilon _{n}}$ determined via (\ref{MS}) by the density operator in $%
\mathcal{N}_{r+\epsilon _{n}}$ that represents $\rho $). Indeed, Dieks'
suggestion appears to be that when we take successively smaller values for $%
\epsilon $ (holding $r$ fixed), and choose a type I interpolating factor at
each stage, we should be getting progressively better approximations to the
set of observables that are truly definite at the origin. What better
candidate for that set can there be than an intersection like $%
\bigcap_{n=0}^{\infty }\mathcal{M}_{r+\epsilon _{n}}$?

Unfortunately, we have no guarantee that \textit{this} intersection, unlike $%
\bigcap_{n=0}^{\infty }\mathcal{N}_{r+\epsilon _{n}}$ itself, is independent
of the particular sequence $\{\epsilon _{n}\}$ or its starting value $%
\epsilon _{0}=\epsilon $. The reason any intersection of form $%
\bigcap_{n=0}^{\infty }\mathcal{N}_{r+\epsilon _{n}}$ is so independent is
because $\mathcal{N}_{r+\epsilon _{n}}\supset \mathcal{N}_{r+\epsilon
_{n+1}} $ for all $n$. But this does \textit{not} imply $\mathcal{M}%
_{r+\epsilon _{n}}\supset \mathcal{M}_{r+\epsilon _{n+1}}$. To take just a
trivial example: when $\rho $ is a pure state of $\mathcal{N}_{r+\epsilon
_{n}}$ that induces a mixed state on the proper subalgebra $\mathcal{N}%
_{r+\epsilon _{n+1}}$of $\mathcal{N}_{r+\epsilon _{n}}$, $\mathcal{M}%
_{r+\epsilon _{n+1}}$ will contain observables with dispersion in the state $%
\rho $, but $\mathcal{M}_{r+\epsilon _{n}}$ will not.

We conclude that there is little prospect of eliminating the arbitrariness
in Dieks' proposal and making it well-defined. One ought to look for
another, \textit{intrinsic} way to pick out the definite-valued observables
in $\mathcal{R}(\diamondsuit _{r})$ that does not depend on special
assumptions such as the split property.

\section{The Modal Interpretation for Arbitrary von Neumann Algebras}

There are two salient features of the algebra $\mathcal{M}_{S}$ in Eq. (\ref
{MS}) that make it an attractive set of definite-valued observables to modal
interpreters. First, $\mathcal{M}_{S}$ is locally determined by the quantum
state $D_{S}$ of system $S$ together with the structure of its algebra of
observables. In particular, there is no need to add any additional structure
to the standard formalism of quantum theory to pick out $S$'s properties.
Second, the restriction of the state $D_{S}$ to the subalgebra $\mathcal{M}%
_{S}$ is a mixture of dispersion-free states (given by the density operators
one obtains by renormalizing the (non-null) spectral projections of $D_{S}$%
). This second feature is what makes it possible to think of the observables
in \thinspace $\mathcal{M}_{S}$ as possessing definite values distributed in
accordance with standard Born rule statistics \cite{clifred}. Let us see,
then, whether we can generalize these two features to come up with a
proposal for the definite-valued observables of a system described by an
arbitrary von Neumann algebra $\mathcal{R}$ (acting on some Hilbert space 
\textsf{H}) in an arbitrary state $\rho $ of $\mathcal{R}$.

Generally, a state $\rho $ of $\mathcal{R}$ will be a mixture of
dispersion-free states on a subalgebra $\mathcal{S\subseteq R}$ just in case
there is a probability measure $\mu _{\rho }$ on the space $\Lambda $ of
dispersion-free states of $\mathcal{S}$ such that 
\begin{equation}
\rho (A)=\int_{\Lambda }\omega _{\lambda }(A)d\mu _{\rho }(\lambda ),\;\text{%
for all }A\in \mathcal{S}\text{,}  \label{beable}
\end{equation}
where $\omega _{\lambda }(A^{2})=\omega _{\lambda }(A)^{2}$ for all
self-adjoint elements $A\in \mathcal{S}$. This somewhat cumbersome condition
turns out to be equivalent (\cite{hc}, Prop. 2.2(ii)) to simply requiring
that 
\begin{equation}
\rho ([A,B]^{*}[A,B])=0\;\text{for all\ }A,B\in \mathcal{S}.  \label{beable2}
\end{equation}
In particular, $\rho $ can always be represented as a mixture of
dispersion-free states on any abelian subalgebra $\mathcal{S\subseteq R}$.
Conversely, if $\rho $ is a faithful state of $\mathcal{R}$, i.e., $\rho $
maps no nonzero positive elements of $\mathcal{R}$ to zero, then the \textit{%
only} subalgebras that allow $\rho $ to be represented as a mixture of
dispersion-free states are the abelian ones. There is now an easy way to
pick out a subalgebra $\mathcal{S\subseteq R}$ with this property, using
only $\rho $ and the algebraic operations available within $\mathcal{R}$.

Consider the following two mathematical objects explicitly defined in terms
of $\mathcal{R}$ and $\rho $. First, the support projection of the state $%
\rho $ in $\mathcal{R}$, defined by

\begin{equation}
P_{\rho ,\mathcal{R}}\equiv {\Large \wedge }\{P=P^{2}=P^{*}\in \mathcal{R}
:\rho (P)=1\},
\end{equation}
which is simply the smallest projection in $\mathcal{R}$ that the state $%
\rho $ `makes true'. Second, there is the centralizer subalgebra of the
state $\rho $ in $\mathcal{R}$, defined by

\begin{equation}
\mathcal{C}_{\rho ,\mathcal{R}}\equiv \{A\in \mathcal{R}:\rho ([A,B])=0\;%
\text{for all }B\in \mathcal{R}\}.
\end{equation}
For any von Neumann algebra $\mathcal{K}$, let $\mathcal{Z}$($\mathcal{K}$)$%
\equiv \mathcal{K}\cap \mathcal{K}^{\prime }$, the center algebra of $%
\mathcal{K}$. Then it is reasonable for the modal interpreter to take as
definite-valued all the observables that lie in the direct sum 
\begin{equation}
\mathcal{S}=\mathcal{M}_{\rho ,\mathcal{R}}\equiv P_{\rho ,\mathcal{R}%
}^{\perp }\mathcal{R}P_{\rho ,\mathcal{R}}^{\perp }+\mathcal{Z}(\mathcal{C}%
_{\rho ,\mathcal{R}})P_{\rho ,\mathcal{R}}\subseteq \mathcal{R},
\label{general}
\end{equation}
where the algebra in the first summand acts on the subspace $P_{\rho ,%
\mathcal{R}}^{\perp }\mathsf{H}$ and that of the second acts on $P_{\rho ,%
\mathcal{R}}\mathsf{H.}$ The state $\rho $ is a mixture of dispersion-free
states on $\mathcal{M}_{\rho ,\mathcal{R}}$, by (\ref{beable2}), because $%
\rho $ maps all elements of the form $P_{\rho ,\mathcal{R}}^{\perp }\mathcal{%
R}P_{\rho ,\mathcal{R}}^{\perp }$ to zero, and the product of the
commutators of any two elements of $\mathcal{Z}(\mathcal{C}_{\rho ,\mathcal{R%
}})P_{\rho ,\mathcal{R}}$ also gets mapped to zero, for the trivial reason
that $\mathcal{Z}(\mathcal{C}_{\rho ,\mathcal{R}})$ is abelian.

The set $\mathcal{M}_{\rho ,\mathcal{R}}$ directly generalizes the algebra
of Eq. (\ref{MS}) to the non-type I case where the algebra of observables of
the system does not contain a density operator representative of the state $%
\rho $. Assuming the type I case, $\mathcal{R}$ $\approx $ $\mathcal{B}(%
\overline{\mathsf{H}})$ for some Hilbert space $\overline{\mathsf{H}}$, $%
\rho $ is given by a density operator $D$ on $\overline{\mathsf{H}}$, $%
P_{\rho ,\mathcal{R}}$ is equivalent to the range projection of $D$, and $%
\mathcal{Z}(\mathcal{C}_{\rho ,\mathcal{R}})\approx \mathcal{Z}(\mathcal{C}%
_{D,\mathcal{B}(\overline{\mathsf{H}})})$. So to show that $\mathcal{M}%
_{\rho ,\mathcal{R}}$ is isomorphic to the algebra of Eq. (\ref{MS}), it
suffices to establish that $\mathcal{Z}(\mathcal{C}_{D,\mathcal{B}(\overline{%
\mathsf{H}})})=$ $D^{\prime \prime }$. It is easy to see that $\mathcal{C}%
_{D,\mathcal{B}(\overline{\mathsf{H}})}=D^{\prime }$ (invoking cyclicity and
positive-definiteness of the trace), thus $\mathcal{Z}(\mathcal{C}_{D,%
\mathcal{B}(\overline{\mathsf{H}})})=$ $D^{\prime }\cap D^{\prime \prime }$.
However, since $D^{\prime }$ always contains a maximal abelian subalgebra of 
$\mathcal{B}(\overline{\mathsf{H}})$ (viz., that generated by the
projections onto any complete orthonormal basis of eigenvectors for $D$), we
always have $D^{\prime \prime }\subseteq D^{\prime }$.

Choosing $\mathcal{M}_{\rho ,\mathcal{R}}$ is certainly not the only way to
pick a subalgebra $\mathcal{S\subseteq R}$ that is definable in terms of $%
\rho $ and $\mathcal{R}$ and allows $\rho $ to be represented as a mixture
of dispersion-free states. There is the obvious orthodox alternative one can
always consider, viz., the definite algebra of $\rho $ in $\mathcal{R}$, 
\begin{equation}
\mathcal{S}=\mathcal{O}_{\rho ,\mathcal{R}}\equiv \{A\in \mathcal{R}:\rho
(AB)=\rho (A)\rho (B)\;\text{for all }B\in \mathcal{R}\},
\end{equation}
which coincides with the complex span of all self-adjoint members of $%
\mathcal{R}$ on which $\rho $ is dispersion-free (\cite{hc}, p. 2445). Note,
however, that we always have $\mathcal{O}_{\rho ,\mathcal{R}}\subseteq 
\mathcal{M}_{\rho ,\mathcal{R}}$. Indeed the problem is that the orthodox
choice $\mathcal{O}_{\rho ,\mathcal{R}}$ generally will contain far too few
definite-valued observables to solve the measurement problem. For example,
when $\rho $ is faithful --- and there will always be a norm dense set of
states of $\mathcal{R}$ that are --- we get just $\mathcal{O}_{\rho ,%
\mathcal{R}}=\frak{C}I$. Thus it is natural for a modal interpreter to
require that the choice of $\mathcal{S\subseteq R}$ be maximal. In the case
where $\rho $ is faithful, we now show that this singles out the choice $%
\mathcal{S}=\mathcal{M}_{\rho ,\mathcal{R}}=\mathcal{Z}(\mathcal{C}_{\rho ,%
\mathcal{R}})$ uniquely (and we conjecture that a similar uniqueness result
holds for the more general expression for $\mathcal{M}_{\rho ,\mathcal{R}}$
in Eq. (\ref{general}), using the fact that an arbitrary state $\rho $
always renormalizes to a faithful state on $P_{\rho ,\mathcal{R}}\mathcal{R}%
P_{\rho ,\mathcal{R}}$).

\mathstrut

\textbf{Proposition 1} \textit{Let }$\mathcal{R}$\textit{\ be a von Neumann
algebra and }$\rho $\textit{\ a faithful normal state of }$\mathcal{R}$%
\textit{\ with centralizer }$\mathcal{C}_{\rho ,\mathcal{R}}\;\mathcal{%
\subseteq R}$\textit{. Then }$\mathcal{Z}(\mathcal{C}_{\rho ,\mathcal{R}})$%
\textit{, the center of }$\mathcal{C}_{\rho ,\mathcal{R}}$\textit{, is the
unique subalgebra }$\mathcal{S\subseteq R}$\textit{\ such that:}

\begin{enumerate}
\item  \textit{The restriction of }$\rho $\textit{\ to }$\mathcal{S}$\textit{%
\ \ is a mixture of dispersion-free states.}

\item  $\mathcal{S}$\textit{\ is definable solely in terms of }$\rho $%
\textit{\ and the algebraic structure of }$\mathcal{R}$\textit{.}

\item  $\mathcal{S}$\textit{\ is maximal with respect to properties 1. and 2.%
}
\end{enumerate}

Proof: By 3., it suffices to show than any $\mathcal{S\subseteq R}$
satisfying 1. and 2. is contained in $\mathcal{Z}(\mathcal{C}_{\rho ,%
\mathcal{R}})$. And for this, it suffices (because von Neumann algebras are
generated by their projections) to show that \underline{$\mathcal{S}$}, the
subset of projections in $\mathcal{S}$, is contained in $\mathcal{Z}(%
\mathcal{C}_{\rho ,\mathcal{R}})$. Recall also that, as a consequence of 1.
and the faithfulness of $\rho $, $\mathcal{S}$ must be abelian. And in
virtue of 2., any automorphism $\,\sigma :$ $\mathcal{R\longrightarrow R}$
that preserves the state $\rho $ in the sense that $\rho \circ \sigma =\rho $%
, must leave the set $\mathcal{S}$ (not necessarily pointwise) invariant,
i.e., $\sigma (\mathcal{S})=\mathcal{S}$.

\underline{$\mathcal{S}$} $\mathcal{\subseteq C}_{\rho ,\mathcal{R}}^{\prime
}$. Any unitary operator $U\in \mathcal{C}_{\rho ,\mathcal{R}}$ defines an
inner automorphism on $\mathcal{R}$ that leaves $\rho $ invariant, therefore 
$U\mathcal{S}U^{-1}=$ $\mathcal{S}$. Since $\mathcal{S}$ is abelian, $%
[UPU^{-1},P]=0$ for each $P\in $ \underline{$\mathcal{S}$} and all unitary $%
U\in \mathcal{C}_{\rho ,\mathcal{R}}$. By Lemma 4.2 of \cite{hc} (with $%
\frak{V}=\mathcal{C}_{\rho ,\mathcal{R}}^{\prime \prime }=\mathcal{C}_{\rho ,%
\mathcal{R}}$), this implies that $P\in \mathcal{C}_{\rho ,\mathcal{R}%
}^{\prime }$.

\underline{$\mathcal{S}$} $\mathcal{\subseteq C}_{\rho ,\mathcal{R}}$. Since 
$\rho $ is faithful, there is a one-parameter group $\{\sigma _{t}:t\in 
\frak{R}\}$ of automorphisms of $\mathcal{R}$ --- the modular automorphism
group of $\mathcal{R}$ determined by $\rho $ (\cite{KR}, Sec. 9.2) ---
leaving $\rho $ invariant. Since $\mathcal{C}_{\rho ,\mathcal{R}}$ consists
precisely of the fixed points of the modular group (\cite{KR}, Prop.
9.2.14), it suffices to show that it leaves the individual elements of 
\underline{$\mathcal{S}$} fixed. For this, we use the fact that the modular
group satisfies the KMS condition with respect to $\rho $: for each $A,B\in 
\mathcal{R}$, there is a complex-valued function $f$, bounded and continuous
on the strip $\{z\in \frak{C}:0\leq \func{Im}z\leq 1\}$ in the complex
plane, and analytic on the interior of that strip, such that 
\begin{equation}
f(t)=\rho (\sigma _{t}(A)B),\;f(t+i)=\rho (B\sigma _{t}(A)),\;t\in \frak{R}.
\end{equation}
In fact, we shall need only one simple consequence of the KMS condition,
viz., if $f(t)=f(t+i)$ for all $t\in \frak{R}$, then $f$ is constant (\cite
{KR}, p. 611).

Fix an arbitrary projection $P\in $ \underline{$\mathcal{S}$}$.$ Since the
modular automorphism group must leave $\mathcal{S}$ as a whole invariant,
and $\mathcal{S}$ is abelian, $[\sigma _{t}(P),P^{\perp }]=0$ for all $t\in 
\frak{R}$. However, there exists a function $f$ with the above properties
such that 
\[
f(t)=\rho (\sigma _{t}(P)P^{\perp })=\rho (P^{\perp }\sigma
_{t}(P))=f(t+i),\;t\in \frak{R}, 
\]
so it follows that $f$ is constant. In particular, since $\sigma
_{0}(P)P^{\perp }=PP^{\perp }=0$, $f$ is identically zero, and $\rho (\sigma
_{t}(P)P^{\perp })=0$ for all $t\in \frak{R}$. And since $\sigma _{t}(P)$
and $P^{\perp }$ are commuting projections, their product is a (positive)
projection, so that the faithfulness of $\rho $ requires that $\sigma
_{t}(P)P^{\perp }=0$, or equivalently $\sigma _{t}(P)=\sigma _{t}(P)P$, for
all $t\in \frak{R}$. Running through the exact same argument, starting with $%
P^{\perp }\in $ \underline{$\mathcal{S}$} in place of $P$, yields $\sigma
_{t}(P^{\perp })=\sigma _{t}(P^{\perp })P^{\perp }$, or equivalently, $%
\sigma _{t}(P)P=P$, for all $t\in \frak{R}$. Together with $\sigma
_{t}(P)=\sigma _{t}(P)P$, this implies that $\sigma _{t}(P)=P$ for all $t\in 
\frak{R}$. $QED.$

\mathstrut

The choice $\mathcal{M}_{\rho ,\mathcal{R}}=\mathcal{Z}(\mathcal{C}_{\rho ,%
\mathcal{R}})$ has another feature that generalizes a natural consequence of
the modal interpretation of nonrelativistic quantum theory. Suppose the
universal state $x\in $\textsf{H}$_{U}$ defines a faithful state $\rho _{x}$
on both $\mathcal{B}(\mathsf{H}_{S})$ and $\mathcal{B}(\mathsf{H}_{\overline{%
S}})$. This requires that $\dim \mathsf{H}_{S}$ $=\dim \mathsf{H}_{S}=n$
(possibly $\infty $), and, furthermore, that any Schmidt decomposition of
the state vector $x$ relative to the factorization \textsf{H}$_{U}=\mathsf{H}%
_{S}\otimes \mathsf{H}_{\overline{S}}$ takes the form 
\begin{equation}
x=\sum_{i=1}^{n}c_{i}v_{i}\otimes w_{i},\;c_{i}\neq 0\;\text{for all\ }i=1%
\text{ to }n\text{,}  \label{herro}
\end{equation}
where the vectors $v_{i}$ and $w_{i}$ are complete orthonormal bases in
their respective spaces. As is well-known, for each distinct eigenvalue $%
\widetilde{\lambda }_{j}$ for $D_{S}$, the span of the vectors $v_{i}$ for
which $|c_{i}|^{2}=\widetilde{\lambda }_{j}$ coincides with the range of the 
$\widetilde{\lambda }_{j}$-eigenprojection of $D_{S}$, and similarly for $D_{%
\overline{S}}$ . Consequently, there is a natural bijective correspondence
between the properties represented by the projections in the two sets $%
\mathcal{M}_{S}$ $=D_{S}^{\prime \prime }$ and $\mathcal{M}_{\overline{S}}$ $%
=D_{\overline{S}}^{\prime \prime }$: any definite property $S$ happens to
possess is strictly correlated to a unique property of its environment $%
\overline{S}$ that occurs with the same frequency. More formally, for any $%
P\in \mathcal{M}_{S}$, there is a unique $\overline{P}\in \mathcal{M}_{%
\overline{S}}$ satisfying 
\begin{equation}
(x,P\overline{P}x)=(x,Px)=(x,\overline{P}x).  \label{hello}
\end{equation}
To see this, note that any $P\in \mathcal{M}_{S}$ (in this case, the set of
all functions of $D_{S}^{{}}$) is a sum of spectral projections of $D_{S}$.
Let $\overline{P}\in \mathcal{M}_{\overline{S}}$ be the sum of the
corresponding spectral projections of $D_{\overline{S}}$ for the same
eigenvalues. Then it is evident from the form of the state expansion in (\ref
{herro}) that $\overline{P}$ has the property in (\ref{hello}), and no other
projection in $\mathcal{M}_{\overline{S}}$ does. This has led some
non-atomic modal interpreters, such as Kochen \cite{kochen}, to interpret
each property $P$ of $S$, not as a property that $S$ possesses absolutely,
but only in relation to its environment $\overline{S}$ possessing the
corresponding property $\overline{P}$.

For a general von Neumann factor $\mathcal{R}$, $(\mathcal{R}\cup \mathcal{R}%
^{\prime })^{\prime \prime }$ $=\mathcal{B}(\mathsf{H})$ need not be
isomorphic to the tensor product $\mathcal{R}\otimes \mathcal{R}^{\prime }$
(particularly when $\mathcal{R}$ is type III, for then $\mathcal{R}\otimes 
\mathcal{R}^{\prime }$ must be type III as well). Therefore, there is no
direct analogue of a Schmidt decomposition for a pure state $x\in \mathsf{H}$
relative to the factorization $(\mathcal{R}\cup \mathcal{R}^{\prime
})^{\prime \prime }$ of $\mathcal{B}(\mathsf{H})$. Nevertheless, we show
next that there is still the same strict correlation between definite
properties in $\mathcal{M}_{\rho _{x},\mathcal{R}}$ $=\mathcal{Z}(\mathcal{C}%
_{\rho _{x},\mathcal{R}})$ and $\mathcal{M}_{\rho _{x},\mathcal{R}^{\prime
}} $ $=\mathcal{Z}(\mathcal{C}_{\rho _{x},\mathcal{R}^{\prime }})$.

\mathstrut

\textbf{Proposition 2} \textit{Let }$\mathcal{R}$\textit{\ be a von Neumann
algebra acting on a Hilbert space }$\mathsf{H}$\textit{, and suppose }$x\in 
\mathsf{H}$\textit{\ induces a state }$\rho _{x}$\textit{\ that is faithful
on both }$\mathcal{R}$\textit{\ and }$\mathcal{R}^{\prime }$\textit{. Then
for any projection }$P\in \mathcal{Z}(\mathcal{C}_{\rho _{x},\mathcal{R}})$%
\textit{, there is a unique projection }$\overline{P}\in \mathcal{Z}(%
\mathcal{C}_{\rho _{x},\mathcal{R}^{\prime }})$\textit{\ such that }$(x,P%
\overline{P}x)=(x,Px)=(x,\overline{P}x)$\textit{.}

Proof: For any fixed $A\in \mathcal{R},$ call an element $B\in \mathcal{R}%
^{\prime }$ a double for $A$ (in state $x$) just in case $Ax=Bx$ and $%
A^{*}x=B^{*}x$ . By an elementary application of modular theory, Werner (%
\cite{werner}, Sec. II) has shown that $\mathcal{C}_{\rho _{x},\mathcal{R}}$
consists precisely of those elements of $\mathcal{R}$ with doubles in $%
\mathcal{R}^{\prime }$ (with respect to $x$). Moreover, the double of any
element of $\mathcal{R}$ clearly has to be unique, by the faithfulness of $%
\rho _{x}$ on $\mathcal{R}^{\prime }$. Now it is easy to see (again using
the faithfulness of $\rho _{x}$) that the double of any projection $P\in 
\mathcal{C}_{\rho _{x},\mathcal{R}}$ is a projection $\overline{P}\in 
\mathcal{C}_{\rho _{x},\mathcal{R}^{\prime }}$ satisfying (\ref{hello}). We
claim that whenever $P\in \mathcal{C}_{x,\mathcal{R}}^{\prime }$, we have $%
\overline{P}\in \mathcal{C}_{x,\mathcal{R}^{\prime }}^{\prime }$. For this,
it suffices to show $P\in \mathcal{C}_{x,\mathcal{R}}^{\prime }$ implies
that for arbitrary $B\in $ $\mathcal{C}_{x,\mathcal{R}^{\prime }}$, $[%
\overline{P},B]x=0$ (and then $[\overline{P},B]$ itself is zero, since $\rho
_{x}$ is faithful). Letting $A\in $ $\mathcal{C}_{x,\mathcal{R}}$ be the
double of $B$ in $\mathcal{R}$, we get 
\begin{equation}
\overline{P}Bx-B\overline{P}x=\overline{P}Ax-BPx=A\overline{P}%
x-PBx=APx-PAx=0,
\end{equation}
as required. Finally, were there another projection $\widetilde{P}\in 
\mathcal{Z}(\mathcal{C}_{\rho _{x},\mathcal{R}^{\prime }})$ satisfying (\ref
{hello}), then by exploiting the fact that $\overline{P}$ is $P$'s double in 
$\mathcal{R}^{\prime }$, we get $(x,\overline{P}\widetilde{P}x)=(x,\overline{%
P}x)=(x,\widetilde{P}x)$; or, equivalently, 
\begin{equation}
(x,\overline{P}\widetilde{P}^{\perp }x)=(x,\overline{P}^{\perp }\widetilde{P}%
x)=0.  \label{hi}
\end{equation}
Since $\mathcal{Z}(\mathcal{C}_{\rho _{x},\mathcal{R}^{\prime }})$ is
abelian, both $\overline{P}\widetilde{P}^{\perp }$ and $\overline{P}^{\perp }%
\widetilde{P}$ are (positive) projections in $\mathcal{R}$. But as $\rho
_{x} $ is faithful on $\mathcal{R}$, Eqs. (\ref{hi}) entail that $\overline{P%
}\widetilde{P}^{\perp }=\overline{P}^{\perp }\widetilde{P}=0$, which in turn
implies that $\overline{P}=\widetilde{P}$, as required for uniqueness. $QED.$

\mathstrut

Let us return now to the problem of picking out a set of definite-valued
observables localized in a diamond region with associated algebra $\mathcal{R%
}(\diamondsuit _{r})$. Let $\rho $ be any pure state of the field that
induces a faithful state on $\mathcal{R}(\diamondsuit _{r})$; for example, $%
\rho $ could be the vacuum or any one of the dense set of states of a field
with bounded energy (by the Reeh-Schlieder theorem --- see \cite{horuzy},
Thm. 1.3.1). By Proposition 1, the definite-valued observables in $\mathcal{R%
}(\diamondsuit _{r})$ are simply those in the subalgebra $\mathcal{Z}(%
\mathcal{C}_{\rho ,\mathcal{R}(\diamondsuit _{r})})$. Note that this
proposal yields observables all of which have an exact spacetime
localization within the open set $\diamondsuit _{r}$ and are picked out
intrinsically by the local algebra $\mathcal{R}(\diamondsuit _{r})$ and the
field state $\rho $. Contrary to Dieks' pessimistic conclusion, we \textit{%
can} take open spacetime regions as fundamental for determining the
definite-valued observables. In fact, this proposal works independent of the
size of $r$, and so could also be embraced by \textit{non}-atomic modal
interpreters not wishing to commit themselves to a particular partition of
the field into subsystems (or to thinking from the outset in terms of
approximately point-localized field observables). Finally, note that since
the algebra of a diamond region $\diamondsuit _{r}$ satisfies duality with
respect to the algebra of its spacelike complement $\diamondsuit
_{r}^{\prime }$, i.e., $\mathcal{R}(\diamondsuit _{r})^{\prime }=\mathcal{R}%
(\diamondsuit _{r}^{\prime })$ (\cite{haag}, p. 145), Proposition 2 tells us
that there is a natural bijective correspondence between the properties in $%
\mathcal{Z}(\mathcal{C}_{\rho ,\mathcal{R}(\diamondsuit _{_{r}})})$ and
strictly correlated properties in $\mathcal{Z}(\mathcal{C}_{\rho ,\mathcal{R}%
(\diamondsuit _{r}^{\prime })})$ associated with the complement region.

\section{A Potential Difficulty with Ergodic States}

We have seen that there \textit{is}, after all, a well-motivated and
unambiguous prescription extending the standard modal interpretation of
nonrelativistic quantum theory to the local algebras of quantum field
theory. We also, now, have a natural standard of comparison with Galilean
quantum field theory. At least in the case of free fields, it is possible to
build up local algebras in $M$ from spatially smeared ``field algebras''
defined on spacelike hyperplanes in $M$. A diamond region corresponds to the
domain of dependence of a spatial region in a hyperplane, and it can be
shown that the algebra of that spatial region will also be type III and,
indeed, \textit{coincide} with its domain of dependence algebra (\cite
{horuzy}, Prop. 3.3.2, Thm. 3.3.4). These type III spatial algebras in $M$,
and the definite-valued observables therein, are what should be compared, in
the nonrelativistic limit, to the corresponding equal time spatial algebras
defined on simultaneity slices of Galilean spacetime. Unfortunately, since
the algebras in the Galilean case are invariably type I (\cite{horuzy}, p.
35), this limit is bound to be mathematically singular, and its \textit{%
physical} characterization needs to be dealt with carefully. But this is a
problem for \textit{any} would-be interpreter of relativistic quantum field
theory, not just modal interpreters. All we should require of them, at this
stage, is that they be able to say something sensible in the relativistic
case about the local observables with definite values (which was, indeed,
Dieks' original goal). However, as we now explain, it is not clear whether
even this goal can be attained.

If $\mathcal{R}$ $\approx $ $\mathcal{B}(\overline{\mathsf{H}})$ is type I,
it possesses at most one faithful state $\rho $ such that $\mathcal{Z}(%
\mathcal{C}_{\rho ,\mathcal{R}})=\frak{C}I$. This is easy to see, because if 
$D\in \mathcal{B}(\overline{\mathsf{H}})$ represents $\rho $, $\mathcal{Z}(%
\mathcal{C}_{\rho ,\mathcal{R}})\approx D^{\prime \prime }$, and $D^{\prime
\prime }=\frak{C}I$ implies that $D$ itself must be a multiple of the
identity. So when $\overline{\mathsf{H}}$ is finite-dimensional, we must
have $D=I/\dim \overline{\mathsf{H}}$, the unique maximally mixed state, and
in the infinite-dimensional case, no such density operator even exists.
Elsewhere Dieks \cite{dieks1} has argued convincingly that there is no
problem when a system, occupying a maximally mixed state, possesses only
trivial properties, because such states are rare and highly unstable under
environmental decoherence (cf. \cite{healeybook}, pp. 99-100). However, the
situation is quite different for the local algebras of algebraic quantum
field theory.

In all physically reasonable models of the axioms of the theory, every local
algebra $\mathcal{R}(O)$ is isomorphic to the unique (up to isomorphism)
hyperfinite type III$_{1}$ factor (\cite{haag}, Sec. V.6; \cite{russians},
Sec. 17.2). In that case, there is a novel way to obtain $\mathcal{Z}(%
\mathcal{C}_{\rho ,\mathcal{R}(O)})=\frak{C}I$, namely, when the state $\rho 
$ of $\mathcal{R}(O)$ is an ergodic state\textit{\ }\cite{tak,barn}, i.e., $%
\rho $ possesses a trivial centralizer in $\mathcal{R}(O)$. (Were $\mathcal{R%
}$ a nonabelian type I factor, this would be impossible, since $D^{\prime }=%
\frak{C}I$ implies $D^{\prime \prime }=\mathcal{B}(\overline{\mathsf{H}}%
)\;\approx $ $\mathcal{R}$, which is patently false.) In fact, we have the
following result.

\mathstrut

\textbf{Proposition 3} \textit{If }$\mathcal{R}$\textit{\ is the hyperfinite
type }III$_{1}$\textit{\ factor, there is a norm dense set of unit vectors
in the Hilbert space }\textsf{H}\textit{\ on which }$\mathcal{R}$\textit{\
acts that induce faithful states on }$\mathcal{R}$\textit{\ with trivial
centralizers (i.e., ergodic states).}

Proof: First recall the following facts provable from the axioms of
algebraic quantum field theory: (i) the vacuum state of a field on $M$ has a
trivial centralizer in the algebra of any Rindler wedge (\cite{russians},
Sec. 16.1.1); (ii) the vacuum is faithful for any wedge algebra (by the
Reeh-Schlieder theorem); and (iii) wedge algebras are hyperfinite type III$%
_{1}$ factors (\cite{russians}, Ex. 16.2.14, pp. 426-7). Since being
faithful and having a trivial centralizer are isomorphic invariants, it
follows that \textit{any} instantiation $\mathcal{R}$ of the hyperfinite
type III$_{1}$ factor possesses at least one faithful normal state $\rho $
with trivial centralizer (even when $\mathcal{R}$ is the algebra of a 
\textit{bounded} open region, like a diamond). Now since $\mathcal{R}$ is
type III, all its states are vector states (combine \cite{sak}, Cor. 2.9.28
with \cite{KR}, Thm. 7.2.3); in particular, $\rho =\rho _{x}$ for some unit
vector $x\in $\textsf{H. }Furthermore, by the homogeneity of the state space
of type III$_{1}$ factors (\cite{connes}, Cor. 6), the set of all unit
vectors of the form $UU^{\prime }x$, with $U\in \mathcal{R}$ and $U^{\prime
}\in \mathcal{R}^{\prime }$ unitary operators, lies dense in \textsf{H}. But
clearly any such vector must again induce a faithful state on $\mathcal{R}$
with trivial centralizer. $QED.$

\mathstrut

Combining Propositions 1 and 3, there will be a whole host of states\textit{%
\ }of any relativistic quantum field in which the modal interpreter is
forced to assert that \textit{no} nontrivial local observables have definite
values! Note, however, that while the set of field states ergodic for any
given type III$_{1}$ local algebra is always dense, this does not
automatically imply that such states are typical or generic. Indeed, results
of Summers and Werner (\cite{sw}, particularly Cor. 2.4) imply that for any
local diamond algebra $\mathcal{R}(\diamondsuit \ss _{r})$, there will 
\textit{also} always be a dense set field states whose centralizers in $%
\mathcal{R}(\diamondsuit _{r})$ contain the hyperfinite type II$_{1}$
factor, and so will \textit{not }be\textit{\ }trivial. Still, the modal
interpreter needs to provide some physical reason for neglecting the densely
many field states that \textit{do} yield trivial definite-valued observables
locally. Obviously instability under decoherence is no longer be relevant.

Perhaps one could try to bypass Proposition 1 by exploiting extra structure
not contained in the particular field state and local algebra to pick out
the definite-valued observables in a region. For example, one might try to
exploit the field's total energy-momentum operator, and, in particular, its
generator of time evolution. In the context of the nonrelativistic modal
interpretation, Bacciagaluppi \textit{et al.} \cite{tough} (cf. also \cite
{BD}, p. 1181) have successfully invoked the analytic properties of the time
evolution of the spectral projections of a system's reduced density operator 
$D_{S}$ to avoid discontinuities that occur in the definite-valued set $%
\mathcal{M}_{S}$ at moments of time where the multiplicity of the
eigenvalues of $D_{S}$ changes. Their methods yield a natural dynamical way,
independent of instability considerations, to avoid the trivial
definite-valued sets determined by maximally mixed density operators. So one
might hope that these same dynamical methods could be extended to type III$%
_{1}$ algebras so as to yield a richer set of properties in a local region
than Proposition 1 allows for ergodic states. In any case, modal
interpreters need to do more work to show that their interpretation yields
sensible local properties in quantum field theory (even before one
considers, with Dieks, how to define Lorentz invariant decoherent histories
of properties).

\bigskip

\begin{center}
{\Large Acknowledgement}
\end{center}

The author wishes to thank Hans Halvorson (for supplying the argument
immediately following Eq. (\ref{hans}) and suggesting the use of the KMS
condition in the proof of Proposition 1), and Reinhard Werner (for helpful
correspondence about centralizers and inspiring Proposition 3).

\end{document}